\documentclass[twocolumn]{aastex62}

\newcommand{\target}{KSP-OT-201611a}
\newcommand{\Ps}{$P_s$}
\newcommand{\Po}{$P_o$}

\newcommand{\ylee}[1]{{\textcolor{black}{#1}}}

\shorttitle{Dwarf nova : KSP-OT-201611a}
\shortauthors{Lee et al.}

\begin{document}

\title{KSP-OT-201611a: A Distant Population II Dwarf Nova Candidate Discovered by the KMTNet Supernova Program
}

\correspondingauthor{Youngdae Lee}
\email{ylee@kasi.re.kr}

\author{Youngdae Lee}
\affil{Korea Astronomy and Space Science Institute, 776, Daedeokdae-ro, Yuseong-gu, Daejeon 34055, Republic of Korea}

\author{Dae-Sik Moon}
\affil{Department of Astronomy and Astrophysics, University of Toronto, 50 St. George Street, Toronto, ON M5S 3H4, Canada}

\author{Sang Chul Kim}
\affil{Korea Astronomy and Space Science Institute, 776, Daedeokdae-ro, Yuseong-gu, Daejeon 34055, Republic of Korea}
\affil{Korea University of Science and Technology (UST), 217 Gajeong-ro, Yuseong-gu, Daejeon 34113, Republic of Korea}

\author{Hong Soo Park}
\affil{Korea Astronomy and Space Science Institute, 776, Daedeokdae-ro, Yuseong-gu, Daejeon 34055, Republic of Korea}

\author{Sang-Mok Cha}
\affil{Korea Astronomy and Space Science Institute, 776, Daedeokdae-ro, Yuseong-gu, Daejeon 34055, Republic of Korea}
\affil{School of Space Research, Kyung Hee University, Yongin 17104, Republic of Korea}

\author{Yongseok Lee}
\affil{Korea Astronomy and Space Science Institute, 776, Daedeokdae-ro, Yuseong-gu, Daejeon 34055, Republic of Korea}
\affil{School of Space Research, Kyung Hee University, Yongin 17104, Republic of Korea}

\begin{abstract}

We present a multi-color, high-cadence photometric study of a distant dwarf nova {\target} discovered by the Korea Microlensing Telescope Network Supernova Program. From October 2016 to May 2017, two outbursts, which comprises a \ylee{super/long} outburst followed by a \ylee{normal/short} outburst separated by $\sim$91 days, were detected in the $BVI$ bands. The shapes and amplitudes of the outbursts reveal the nature of {\target} to be an SU UMa- \ylee{or U Gem-}type dwarf nova. \ylee{Color variations of periodic humps in the super/long outbutst possibly indicate that {\target} is an SU UMa-type dwarf nova.} The super and normal outbursts show distinctively different color evolutions during the outbursts due most likely to the difference \ylee{of time when the cooling wave is formed in} the accretion disk. The outburst peak magnitudes and the orbital period of the dwarf nova indicate that it is at a large Galactocentric distance ($\sim$13.8 kpc) and height ($\sim$1.7 kpc) from the Galactic plane. {\target}, therefore, may provide a rare opportunity to study the accretion disk process of Population II dwarf novae.

 
\end{abstract}


\section{Introduction} \label{sec:intro}

Dwarf novae as a subset of cataclysmic variables are known as close binary systems comprising a white dwarf primary with an accretion disk and a main sequence secondary \citep{War95}. Dwarf novae, in general, have three major types: Z Cam, U Gem, and SU UMa \citep{Osa96} depending on unique characteristics of their light curves. Z Cam types have intervals of constant brightness called standstills, whereas U Gem types are featured by regular quasi-periodic outbursts. SU UMa types, which are more frequently found than the other types \citep{Otu16}, show a combination of normal and superoutbursts. The superoutbursts of SU UMa types typically have longer outburst duration ($\sim$12--20 days) than normal outbursts ($\lesssim$8 days), and the former are usually brighter than the latter by $\sim$0.8 mag in the $V$ band \citep{Pat11,Otu16}. 
Their peak brightness is known to have a tight correlation with the orbital period \citep{War87,Pat11}.

Studies of dwarf novae have been focused on the analysis of observed light curves since they can provide important information about the nature of the observed dwarf novae. \ylee{For eclipsing dwarf novae, the origin of the accretion disk instability leading to outbursts can be investigated based on the shape of the light curves \citep{Vog83,Ioa99,Web99,Bap00}. In this case, outside-in outbursts show symmetric light curves, whereas inside-out outbursts produce asymmetric light curves. Also, if dwarf novae show a delay between the UV and visible light, large and small UV delays mean outside-in and inside-out outbursts, respectively \citep{Can86,Can87}.} The accretion rate can also be investigated using the time interval between outbursts \citep{Sma84b,Sch03}. In addition, if multi-color information is available, more detailed analysis of accretion process and related physical parameters can be conducted, including size, temperature, surface density, and viscosity \citep{May80,Can87}. Unfortunately, however, multi-color observations have been very sparse in dwarf nova observations, limiting our understanding of their origin \citep{Bro18,Shu18}.

The typical $V$-band absolute magnitudes of dwarf novae are within a range of 7--9 mag \citep{War87}, and they have been mostly found in the solar neighborhood ($\lesssim$$1$ kpc) \citep{Dow01,Ozd15}. These nearby dwarf novae are part of the thin disk of the Milky Way \citep{Ak13}. Their mass donors, i.e., the main sequence secondaries, have relatively high metallicity and low velocity \citep{Ak13,Har15,Har16}. Most of the dwarf novae that have been studied so far, therefore, belong to Population I (PopI) group. The properties of low metallicity Population II (PopII) dwarf novae have been poorly studied, mostly due to the lack of observed samples \citep{Haw87,How90,Edm03}. \ylee{Handful of} PopII dwarf novae that have been studied so far, to the best of our knowledge, are those in the globular cluster 47 Tucanae \citep{Edm03} at 4.5 kpc from the Sun \citep{Zoc01} with ${\rm [Fe/H]}$ $\simeq$ $-$0.78 \citep{Thy14}. \ylee{Furthermore, few dwarf novae are spectroscopically classified as thick disk components \citep{Ak13}. The only spectroscopically confirmed PopII dwarf nova is SDSS\,J1507+52 of [Fe/H] $= -1.2$ and $d = 250$\,pc \citep{Uth11}.} Due to the low metallicity, the accretion disk luminosity of PopII dwarf novae is expected to be higher than those of PopI \citep{Ste97}. Surprisingly, however, the dwarf novae in 47 Tucanae have been observed with much smaller accretion disk luminosity than theoretically expected \citep{Edm03}. The origin of this discrepancy is not well understood, and it is imperative to obtain more observational samples of PopII dwarf novae.

In this paper, we present the discovery and high-cadence, multi-color monitoring of a new dwarf nova {\target} most likely belong to PopII group at the distance of $\sim$13.8 kpc from the Galactic center and the height of $\sim$1.7 kpc from the Galactic plane. Section~\ref{sec:discovery} and \ref{sec:lc} provide our discovery and photometry of the source and analysis of its light curves and colors, respectively. We determine the nature of {\target} to be an SU UMa-type dwarf nova and measure orbital period and related parameters in Section~\ref{sec:target}. We discuss the properties of {\target} in Section~\ref{sec:discuss} and provide summary and conclusion in Section~\ref{sec:conclusion}.

\section{{\target} : Discovery and Photometry} \label{sec:discovery}

As part of the Korea Microlensing Telescope Network \citep[KMTNet;][]{Kim16} Supernova Program  \citep[KSP;][]{Moo16}, we have monitored a 4-deg$^2$ area toward the lenticular galaxy NGC 2292 since October 2016. In the program, we have obtained 60-s exposure of $BVI$ bands with typical of $\sim$8-hour cadence. We discovered a new transient source, which we name {\target}, at $(\alpha,\delta)_{\rm J2000}=(06^h\,42^m\,02.1^s, -26^\circ\,00'\,21.6'')$ or $(l,b)=(235.51135^{\circ},-13.45849^{\circ})$ on November 29, 2016 (${\rm MJD} = 57721.17$ day) at a $V$-band magnitude of $19.19 \pm 0.02$ mag (see below for the photometric calibration). 

\begin{figure*}
\begin{center}
\epsscale{1.1}
\plotone{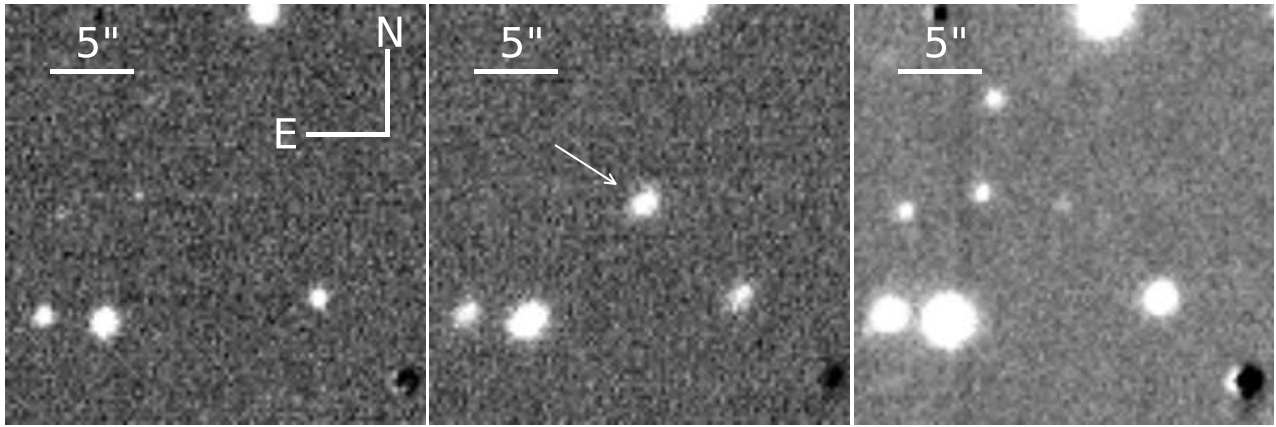}
\caption{$V$-band images before the \ylee{outburst} (left) and at maximum luminosity (middle), and a 170 stacked $V$-band image in quiescence (right). The arrow points to {\target}. \label{fig01}}
\end{center}
\end{figure*}

Figure~\ref{fig01} compares $V$-band images centered on {\target} obtained before the first detection and when the source is at its peak brightness. It also shows a deep image made by stacking 170 individual exposures of 60 seconds obtained when the source was fainter than the detection limit in each image. As seen in the right panel of Figure~\ref{fig01}, {\target} appears as a very faint source of $V = 23.45 \pm 0.11$ mag on the deep stack image (see Section~\ref{sec:quiescent}).

We conducted PSF photometry using DAOPHOT \citep{Ste87} and used the AAVSO Photometric All-Sky Survey database (APASS)\footnote{\url{https://www.aavso.org/apass}} $BVi'$ standard stars for photometric calibration. We used about 300 stars near the source to obtain Penny function based PSF of each image. The typical limiting magnitude of 60-s exposure image obtained in this way is $\sim$$21.5$ mag at $S/N=5$ under 1.2$\arcsec$ seeing in the $V$ band. Given the discrepancy between our $I$-band observations and $i'$-band standard photometric system used in APASS, we adopt the relation $I = i' - 0.4$ known between the KMTNet $I$ band and the APASS $i'$ band \citep[Park et al. in preparation;][]{Par17}. We use the extinction of $A_B = 0.41$ mag, $A_V = 0.32$ mag, and $A_I = 0.20$ mag using the extinction $E(B-V)=0.101$ mag obtained toward \target\ in the Galactic extinction model of \citet{Sch98} with $R_V = 3.1$. \ylee{The extinction-corrected photometric results are shown in Table~\ref{tab01}}. In this paper, we use extinction-corrected magnitudes and colors unless otherwise specified.

\begin{table*} [h!]
\centering
\caption{Photometric catalog for {\target}} \label{tab01}
\renewcommand{\arraystretch}{0.8}
\begin{tabular}{c c c | c c c | c c c}

\hline
\hline
\multicolumn{3}{c|}{$B$} & \multicolumn{3}{c|}{$V$}  & \multicolumn{3}{c}{$I$}           \\
  MJD       &  Mag  &  Error     &    MJD        &   Mag &  Error     &     MJD       & Mag   &  Error      \\
  \[[days]    & [mag] &  [mag]     &    [days]     & [mag] &  [mag]     &     [days]          & [mag] &  [mag]      \\
\hline
57721.16 & 19.27 & 0.02  & 	57721.16 & 19.18 & 0.02  & 	57721.12 & 19.44 & 0.07   \\
57721.27 & 19.34 & 0.02  & 	57721.17 & 19.16 & 0.02  & 	57721.17 & 19.25 & 0.04   \\
57721.31 & 19.35 & 0.02  & 	57721.22 & 19.19 & 0.02  & 	57721.17 & 19.30 & 0.04   \\
57722.22 & 19.22 & 0.01  & 	57721.27 & 19.25 & 0.02  & 	57721.22 & 19.30 & 0.04   \\
57722.31 & 19.00 & 0.01  & 	57721.32 & 19.29 & 0.02  & 	57721.27 & 19.25 & 0.04   \\
57723.18 & 18.99 & 0.01  & 	57722.17 & 18.95 & 0.02  & 	57721.32 & 19.31 & 0.03   \\
57723.22 & 19.09 & 0.01  & 	57722.22 & 19.04 & 0.01  & 	57722.17 & 19.05 & 0.04   \\
57723.27 & 19.00 & 0.01  & 	57722.27 & 19.06 & 0.02  & 	57722.22 & 19.09 & 0.03   \\
57723.32 & 19.05 & 0.01  & 	57722.32 & 18.93 & 0.01  & 	57722.27 & 19.11 & 0.04   \\
57723.59 & 19.16 & 0.03  & 	57723.18 & 18.79 & 0.01  & 	57722.32 & 19.06 & 0.03   \\
57723.64 & 19.14 & 0.03  & 	57723.23 & 18.99 & 0.01  & 	57723.18 & 18.91 & 0.03   \\
57724.18 & 19.18 & 0.01  & 	57723.27 & 18.95 & 0.02  & 	57723.23 & 19.07 & 0.03   \\
57724.23 & 19.15 & 0.01  & 	57723.32 & 18.95 & 0.01  & 	57723.27 & 19.03 & 0.03   \\
57724.28 & 19.04 & 0.01  & 	57723.59 & 19.04 & 0.02  & 	57723.32 & 19.03 & 0.03   \\
57724.33 & 19.21 & 0.01  & 	57723.64 & 18.98 & 0.03  & 	57723.59 & 19.00 & 0.03   \\
57724.67 & 19.22 & 0.03  & 	57723.69 & 18.93 & 0.02  & 	57723.64 & 18.98 & 0.04   \\
57725.18 & 19.25 & 0.02  & 	57724.18 & 19.07 & 0.02  & 	57723.69 & 19.04 & 0.04   \\
57725.23 & 19.17 & 0.01  & 	57724.23 & 19.07 & 0.02  & 	57724.18 & 19.16 & 0.03   \\
57725.28 & 19.29 & 0.01  & 	57724.28 & 18.91 & 0.01  & 	57724.23 & 19.14 & 0.03   \\
57725.32 & 19.23 & 0.01  & 	57724.33 & 19.09 & 0.02  & 	57724.28 & 18.96 & 0.03   \\
57727.18 & 19.50 & 0.02  & 	57724.67 & 19.10 & 0.03  & 	57724.33 & 19.22 & 0.03   \\
57727.23 & 19.50 & 0.02  & 	57725.18 & 19.20 & 0.02  & 	57724.67 & 19.14 & 0.04   \\
57727.28 & 19.55 & 0.02  & 	57725.23 & 19.05 & 0.02  & 	57724.72 & 19.05 & 0.04   \\
57727.33 & 19.53 & 0.02  & 	57725.28 & 19.18 & 0.02  & 	57725.19 & 19.22 & 0.04   \\
57728.19 & 19.67 & 0.02  & 	57725.32 & 19.14 & 0.02  & 	57725.23 & 19.16 & 0.03   \\
57728.24 & 19.64 & 0.02  & 	57727.18 & 19.45 & 0.03  & 	57725.28 & 19.22 & 0.04   \\
57728.29 & 19.59 & 0.02  & 	57727.23 & 19.41 & 0.02  & 	57725.32 & 19.24 & 0.03   \\
57728.34 & 19.69 & 0.02  & 	57727.28 & 19.36 & 0.02  & 	57727.18 & 19.43 & 0.05   \\
57729.11 & 19.84 & 0.04  & 	57727.33 & 19.36 & 0.02  & 	57727.24 & 19.46 & 0.04   \\
57729.16 & 19.64 & 0.02  & 	57728.19 & 19.54 & 0.02  & 	57727.29 & 19.46 & 0.04   \\
57729.21 & 19.81 & 0.02  & 	57728.24 & 19.53 & 0.02  & 	57727.34 & 19.41 & 0.04   \\
57729.26 & 19.70 & 0.02  & 	57728.29 & 19.43 & 0.02  & 	57728.19 & 19.62 & 0.05   \\
57729.31 & 19.72 & 0.01  & 	57728.34 & 19.50 & 0.02  & 	57728.24 & 19.58 & 0.04   \\
57730.21 & 19.84 & 0.04  & 	57729.11 & 19.61 & 0.04  & 	57728.29 & 19.49 & 0.04   \\
57734.28 & 20.68 & 0.11  & 	57729.17 & 19.43 & 0.03  & 	57728.34 & 19.58 & 0.04   \\
57813.79 & 20.36 & 0.04  & 	57729.21 & 19.60 & 0.02  & 	57729.12 & 19.66 & 0.06   \\
57813.83 & 20.06 & 0.03  & 	57729.26 & 19.59 & 0.02  & 	57729.17 & 19.53 & 0.04   \\
57815.05 & 20.33 & 0.04  & 	57729.31 & 19.57 & 0.02  & 	57729.21 & 19.74 & 0.06   \\
57815.10 & 20.35 & 0.04  & 	57729.71 & 19.49 & 0.05  & 	57729.26 & 19.69 & 0.05   \\
57815.79 & 21.02 & 0.10  & 	57730.26 & 19.58 & 0.06  & 	57729.31 & 19.70 & 0.05   \\
57815.84 & 20.90 & 0.09  & 	57734.28 & 20.55 & 0.11  & 	57729.71 & 19.61 & 0.06   \\
         &       &       & 	57734.33 & 20.61 & 0.08  & 	57730.26 & 19.59 & 0.06   \\
         &       &       & 	57813.79 & 20.11 & 0.04  & 	57730.30 & 19.68 & 0.07   \\
         &       &       & 	57813.83 & 20.02 & 0.03  & 	57734.29 & 20.53 & 0.11   \\
         &       &       & 	57815.05 & 20.28 & 0.04  & 	57734.33 & 20.32 & 0.10   \\
         &       &       & 	57815.10 & 20.21 & 0.05  & 	57813.79 & 20.21 & 0.06   \\
         &       &       & 	57815.15 & 20.33 & 0.05  & 	57813.84 & 20.08 & 0.05   \\
         &       &       & 	57815.79 & 21.02 & 0.11  & 	57815.05 & 20.21 & 0.07   \\
         &       &       & 	57815.84 & 20.90 & 0.11  & 	57815.10 & 20.26 & 0.07   \\
         &       &       & 	57815.89 & 20.79 & 0.11  & 	57815.14 & 20.26 & 0.08   \\
         &       &       & 	         &       &       & 	57815.79 & 20.68 & 0.09   \\
         &       &       & 	         &       &       & 	57815.84 & 20.90 & 0.14   \\
\hline
\multicolumn{9}{l}{NOTE : Measured magnitudes are extiction-corrected.} \\

\end{tabular}
\end{table*}

\section{Light Curves and color analysis} \label{sec:lc}

\subsection{Two outbursts} \label{sec:first}

\begin{figure}
\epsscale{1.25}
\plotone{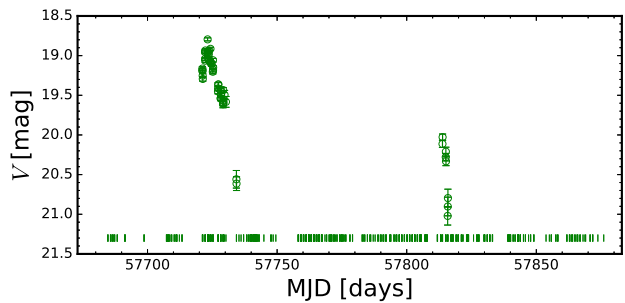}
\caption{The $V$-band light curve of {\target}. Small vertical bars at the bottom of the panel present the $V$-band epochs when observations are made.
\label{fig02}}
\end{figure}

\begin{figure}
\epsscale{1.25}
\plotone{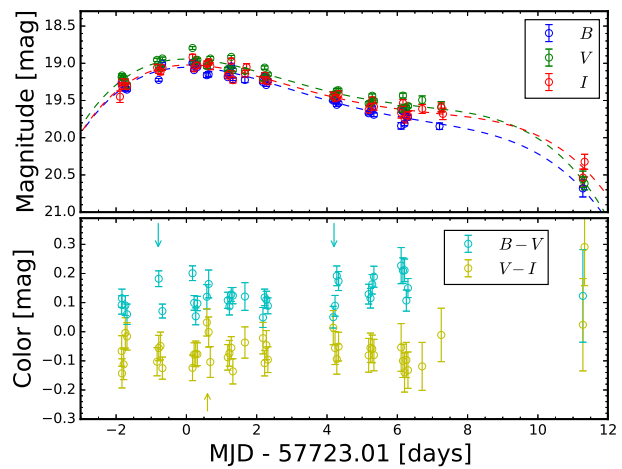}
\caption{(Top) $BVI$ light curves of the first outburst. Dashed lines are fourth-order polynomials for each light curve. (Bottom) Color evolutions of the first outburst. Arrows indicate the locations of notable scatter in colors. \label{fig03}}
\end{figure}

Figure~\ref{fig02} shows the $V$-band light curve of {\target} obtained in a period of 192 days. We detected two outbursts separated by $\sim$91 days: the first, which is brighter, one at ${\rm MJD} \sim 57723$ day  and the second one at ${\rm MJD} \sim 57814$ day. Figure~\ref{fig03} shows $BVI$ light curves and color evolutions of the first outburst. As in the figure, the outburst lasts about 18 days reaching its peak brightness ($V = 18.94$ mag; see below) about two days after the first detection. The light curves overall are an asymmetric shape. After reaching the peak brightness, they appear to enter a relatively flat plateau phase and gradually decay for 10 days, followed by a decline phase below the detection limit.

In order to characterize the light curve, we conduct fourth-order polynomial fits to the light curves as shown by dashed lines in Figure~\ref{fig03}. Based on the fits, we obtain the epochs and magnitudes of the peak as well as the rising rate, decay rate of the plateau, and decline rate. The $V$-band peak magnitude and epoch at the peak brightness are $18.94 \pm0.06$ mag and ${\rm MJD} = 57723.01$ day, respectively. Table~\ref{tab02} contains vital physical parameters of the first outburst obtained from the polynomial fits, including the epoch and magnitude of the peak brightness, rising rate, decay rate of the plateau, and decline rate. It also includes the amplitude and duration of the outburst measured from quiescence magnitude (see Section~\ref{sec:quiescent}).

The bottom panel of Figure~\ref{fig03} shows color evolutions of the first outburst in $B-V$ and $V-I$. The mean colors during the outburst are 0.13 mag ($B-V$) and $-$0.07 ($V-I$), respectively and show no apparent variation as the outburst brightness changes. Although the colors overall mostly constant during the outburst, we do note that there exists notable scatter in colors even for very similar epochs next to each other as indicated by arrows. This scatter in colors is \ylee{probably related to the periodic humps of dwarf novae} (see more details in Section~\ref{ssec:hump}).

\begin{table*} 
\renewcommand{\thetable}{\arabic{table}}
\centering
\caption{Properties of the first outburst} \label{tab02}
\begin{tabular}{c c c c}
\tablewidth{0pt}
\hline
\hline
Parameter &  & Values &  \\
 & $B$ & $V$ & $I$ \\
\hline
\decimals
Peak Magnitude ([mag]) &  19.05 $\pm$ 0.06 &   18.94 $\pm$ 0.06 &  19.02 $\pm$ 0.06 \\
Epoch of Peak brightness ($t_{p_1}$ [day]) &  57723.06 $\pm$ 0.12   &  57723.01 $\pm$ 0.08   & 57723.14 $\pm$ 0.07  \\
Rising rate ($\tau_{r_1}$ [day mag$^{-1}$])   &  0.66 $\pm$ 0.07     & 0.62 $\pm$ 0.07 & 0.85 $\pm$ 0.05     \\
Decay Rate of Plateau ($\tau_{p}$ [day mag$^{-1}$]) &  8.45 $\pm$ 1.93   &  9.36 $\pm$ 0.69  & 10.22 $\pm$ 0.80     \\
Decline Rate ($\tau_{d_1}$ [day mag$^{-1}$]) &  1.00 $\pm$ 0.38   &   0.78 $\pm$ 0.15  & 1.33 $\pm$ 0.18  \\
Amplitude of Outburst ($A_{p_1}$ [mag]) &  4.46 $\pm$ 0.13   &   4.51 $\pm$ 0.13  & 3.66 $\pm$ 0.11  \\
Duration of Outburst ($D_{p_1}$ [day]) & 19.09 $\pm$ 1.65 & 18.43 $\pm$ 0.72 & 19.28 $\pm$ 0.63 \\

\hline

\multicolumn{4}{l}{NOTE : All uncertainties are measured from the bootstrap re-sampling method,} \\
\multicolumn{4}{l}{ but uncertainties of amplitudes are propagated from the peak and quiescence errors.} \\
\end{tabular}
\end{table*}

\begin{figure}
\epsscale{1.25}
\plotone{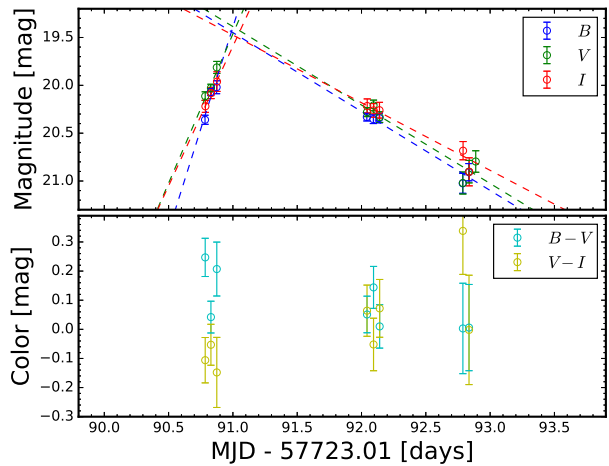}
\caption{(Top) $BVI$ light curves of the second outburst. Dashed lines are first-order polynomials for each light curve at rising and decline phases. (Bottom) Color evolutions of the second outburst. \label{fig04}}
\end{figure}

Figure~\ref{fig04} shows the light curves and color evolutions of the second outburst detected about 91 days after the first outburst. The second outburst is distinctively different from the first outburst in its shape, duration and color evolution. While the first outburst lasts about 18 days with multiple phases (such as rise, plateau, and decline), the second outburst lasts for a much shorter duration of time with much simpler but sharp variation in light curves. \ylee{The lack of the number of observed epochs, which is three separate epochs with eight pointings for each $BVI$ band, makes it difficult to analyze the observed light curves from the second outburst. We, therefore, simply adopted a linear fit to characterize the light curves.} We conduct first-order polynomial fits (dashed lines shown in the top panel of Figure~\ref{fig04}) to the rising and decline parts, separately. Based on the fits, the peak epochs ($57814.03 \pm 0.06$ day) and magnitudes ($19.40 \pm 0.14$ mag in the $V$ band) are calculated from a crossing point between the best-fit models of the rising and decline parts. Inverse slopes of the best-fit models give rising and decline rates. The faster rising rate (0.32 day mag$^{-1}$ in the $V$ band) than the declining rate (1.22 day mag$^{-1}$ in the $V$ band) means that the second outburst is an asymmetric shape. The calculated parameters are shown in Table~\ref{tab03} which also includes amplitude and duration of the second outburst from quiescence (see Section~\ref{sec:quiescent}). When comparing the first and second outbursts, we conclude that the peak magnitude and duration of the second outburst are 0.46 mag fainter and $\sim$3 times shorter than those of the first outburst, respectively.

In the bottom panel of Figure~\ref{fig04}, while $B-V$ color becomes bluer as the outburst progress, $V-I$ color becomes redder. These are clearly different from the color evolution of the first outburst (Figure~\ref{fig03}) where no apparent color evolution is identified. Based on the light curve properties and color evolution pattern, the two outbursts appear to have a different origin (see Section~\ref{disc:outburst}).

\begin{table*} 
\renewcommand{\thetable}{\arabic{table}}
\centering
\caption{Properties of the second outburst} \label{tab03}
\begin{tabular}{c c c c}
\tablewidth{0pt}
\hline
\hline
Parameter &  & Values &  \\
 & $B$ & $V$ & $I$ \\
\hline
\decimals
Peak Magnitude ([mag]) &  19.44 $\pm$ 0.31   &   19.40 $\pm$ 0.14  & 19.50 $\pm$ 0.09  \\
Epoch of Peak brightness ($t_{p_2}$ [day]) &  57814.00 $\pm$ 0.13   &   57814.03 $\pm$ 0.06  & 57814.04 $\pm$ 0.03  \\
Rising Rate ($\tau_{r_2}$ [day mag$^{-1}$]   &  0.23 $\pm$ 0.02    & 0.32 $\pm$ 0.05 & 0.34 $\pm$ 0.02     \\
Decline Rate ($\tau_{d_2}$ [day mag$^{-1}$]) &  1.21 $\pm$ 0.12   &   1.22 $\pm$ 0.17  & 1.41 $\pm$ 0.24  \\
Amplitude of Outburst ($A_{p_2}$ [mag]) &  4.07 $\pm$ 0.25   &   4.05 $\pm$ 0.18  & 3.18 $\pm$ 0.13  \\
Duration of Outburst ($D_{p_2}$ [day]) & 5.90 $\pm$ 1.41 & 6.23 $\pm$ 0.90 & 5.60 $\pm$ 0.51   \\

\hline

\multicolumn{4}{l}{NOTE.-- All uncertainties are measured from the bootstrap re-sampling method,} \\
\multicolumn{4}{l}{ but uncertainties of amplitudes are propagated from peak and quiescence errors.} \\

\end{tabular}
\end{table*}

\subsection{Quiescent phase} \label{sec:quiescent}

\begin{figure}
\epsscale{1.25}
\plotone{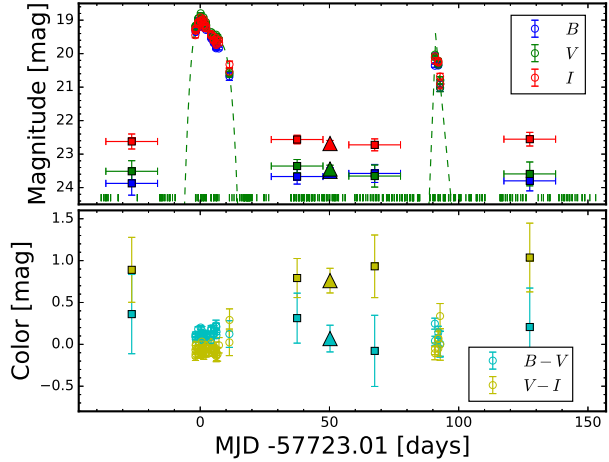}
\caption{$BVI$ light curves (top) and color evolutions (bottom) including the outburst and quiescent phases. Open circles represent outbursts. Quiescent data from 170 stacked images and 40-70 stacked images are depicted as filled triangles and squares, respectively. Horizontal bars at square symbols represent stacking ranges of 20 days. (Top) Blue, green, and red symbols indicate $B$-, $V$-, and $I$-bands, respectively. Small vertical bars at the bottom of the panel represent the $V$-band epochs when the observations are made. Green dashed lines are $V$-band best-polynomial fits. (Bottom) Cyan and yellow symbols represent $B-V$ and $V-I$ colors, respectively. \label{fig05}}
\end{figure}

As we have already shown above in Figure~\ref{fig01}, we stack 170 individual images taken when the source is outside the outbursts to obtain a deep image with much deep detection limit. The source appears at $23.51 \pm 0.11$ mag ($B$), $23.45 \pm 0.11$ ($V$) mag, and $22.68 \pm 0.09$ mag ($I$) on these stack images. In order to see if there is any variation in the source magnitude outside the outbursts, we obtain four stack images at different time intervals: one before the first outburst, two between the two outbursts and one after the second outburst. We use about 40--70 images on average to create each of these deep stack images. The source brightness remains constant in all these four stack images as in Figure~\ref{fig05}. Mean magnitudes of the source in these four stack images are consistent with the magnitudes of the source in the 170 stacked images. We, therefore, adopt the magnitude from the 170 stacked images as the quiescent magnitude of the source.

Figure~\ref{fig05} shows the light curves of the two outbursts (open circles) together with their quiescent magnitudes obtained during the four different time intervals (square symbols) and from the 170 stacked images (triangle symbols). Using the quiescent magnitudes and the polynomial fits obtained in the first and second outbursts, we estimate the duration of the two outbursts to be $18.43 \pm 0.72$ and $6.23 \pm 0.90$ days, respectively (Table~\ref{tab02} and~\ref{tab03}). The differences of the quiescent magnitude and the peak magnitudes in the $V$ band provide amplitudes (4.51 mag for the first outburst and 4.05 mag for the second outburst). In the bottom panel of Figure~\ref{fig05}, The quiescent colors of {\target} are $\sim0.07 \pm 0.16$ mag ($B-V$) and $0.76 \pm 0.15$ mag ($V-I$). The $B-V$ color is similar to the mean $B-V$ color of both the outbursts, while its $V-I$ color is redder than the mean $V-I$ colors of both the outbursts.

\subsection{Color-magnitude and color-color diagrams} \label{sec:cmd}

\begin{figure}
\epsscale{1.25}
\plotone{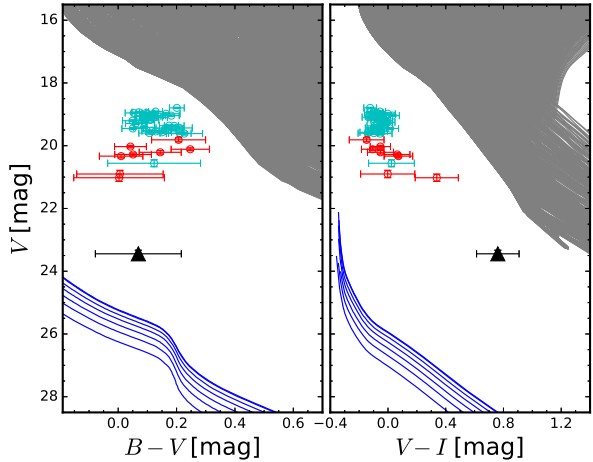}
\caption{ 
Color-magnitude diagrams for the outbursts and quiescent phase. Left and right panels show for $B-V$ and $V-I$ color, respectively. Open cyan/red circles present the first/second outbursts. Black triangles are the quiescent phases from the 170 stacked images. Gray shades are stellar evolution tracks including main sequence and red giant stars with masses of 0.4--5.0 $M_{\odot}$, ages of 0.001--20.0 Gyr, and metallicities ([Fe/H]) of $-3.29$--$0.78$. Blue lines are white dwarf cooling tracks with He+H envelopes, masses of 0.54--1.0 $M_{\odot}$, and ages of 0.0005--14.2 Gyr.
\label{fig06}}

\end{figure}

Figure~\ref{fig06} shows the first/second (open cyan/red circles) outbursts and the quiescent phase (black triangle) in the color-magnitude diagrams. We compare the outbursts and the quiescence with stars in these color-magnitude diagrams. For the stars, we adopt stellar evolutionary tracks from the main sequence to red giant stars \citep{Yi01} and white dwarf cooling tracks \citep{Liu12}. In order to convert the absolute magnitudes of models to apparent magnitudes, we use a distance modulus of 14.33 mag (see Section~\ref{ssec:pars}). The outbursts and quiescence are located between the main sequence (gray shade) and white dwarf cooling sequence (blue lines). This seems to mean that {\target} is not a single star like as a white dwarf or a main sequence star.

\begin{figure}
\epsscale{1.25}
\plotone{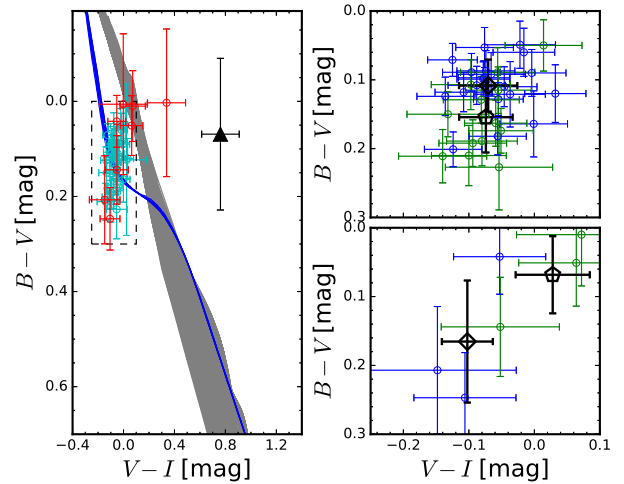}
\caption{(Left) Color-color diagrams including the first/second (open cyan/red circles) outbursts and the quiescent phase (black triangle from 170 stacked images). The gray shade is stellar evolutionary tracks from the main sequence to red giant stars which are the same in Figure~\ref{fig06}. (Right top and bottom) The dashed box in the left panel is focussed. Blue circles are colors around the outburst peak and black diamonds present their mean color. Green circles show colors after the outburst peak and black pentagons are their mean color. Right top and right bottom panels are for the first and the second outbursts, respectively. \label{fig07}}
\end{figure}

The two outbursts and quiescent phase are investigated in a color-color diagram (Figure~\ref{fig07}). The first/second (open cyan/red circles) outbursts present blue colors in both of $B-V$ and $V-I$ colors and cross the white dwarf cooling tracks with an effective temperature of $\sim$13000 K. As seen in the dashed box in the left panel of Figure~\ref{fig07}, the $B-V$ color ranges of the two outbursts are similar and the $V-I$ color ranges are also similar except the reddest $V-I$ color (0.34 mag) for the second outburst. The right two panels focus on the dashed box to show evolutions of color-color changes for the two outbursts. In the two right panels, blue circles are colors around the outburst peak and black diamonds present their mean color. Green circles show colors after the outburst peak and black pentagons are their mean color. For the first outburst (right top panel), the mean colors between around and after the outburst peak are similar within uncertainties, which means that there are no color variations in the first outburst. The mean colors for the second outburst (right bottom panel), however, are clearly different. During the second outburst, $B-V$ and $V-I$ colors become bluer and redder, respectively.




The quiescent phase in the color-color diagram (left panel of Figure~\ref{fig07}) is deviated from the outbursts, especially in $V-I$ color. Since $V-I$ color depends on the emission of cool components, it seems that the processes from the quiescence to the outburst are quite sensitive to the cool components of {\target}.



\section{The nature and parameters of {\target}} \label{sec:target}

\subsection{The nature of {\target}} \label{ssec:nature}

\begin{figure}
\epsscale{1.25}
\plotone{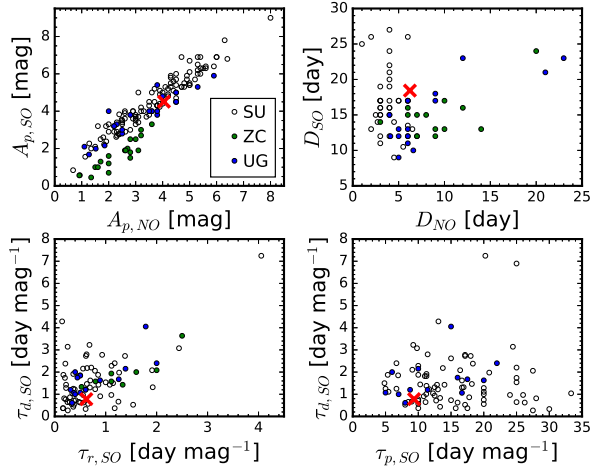}
\caption{The properties of light curves for SU UMa (SU; grey circles),  Z Cam (ZC; green circles), and U Gem (UG; blue circles) types. Red crosses indicate {\target}. Each panel shows combinations of the properties including amplitude ($A_{p,SO}$), duration ($D_{SO}$), rising rate ($\tau_{r,SO}$), decline rate ($\tau_{d,SO}$), and decay rate of the plateau ($\tau_{p,SO}$) for super/long outbursts. They also include the amplitude ($A_{p,NO}$) and duration ($D_{NO}$) for normal/short outbursts. \label{fig08}}
\end{figure}

The properties of a light curve are used to classify the nature of a transient, which include amplitude, duration, rising rate, decline rate, and so on. The $V$-band multiple outbursts with the amplitudes of 4.51 mag and 4.05 mag and duration of 18.43 day and 6.23 day in {\target} are consistent with typical properties of dwarf novae \citep{Osa96,Otu16}. Since there are three types of dwarf novae (SU UMa, Z Cam, and U Gem), we compare the properties of {\target} and the three types of dwarf novae obtained from \citet{Otu16} in Figure~\ref{fig08}. Because the first outburst has larger amplitude and longer duration than the second outburst, we assume that the first and the second outbursts of {\target} are the super/long outburst and normal/short outburst, respectively. In all panels of Figure~\ref{fig08}, the properties of {\target} are well matched with those of SU UMa \ylee{and U Gem types}. This suggests that {\target} is a SU UMa- \ylee{or U Gem-}type dwarf nova. \ylee{Therefore, we expect {\target} to have periodic humps in the super/long outbursts.}


\begin{figure}
\epsscale{1.25}
\plotone{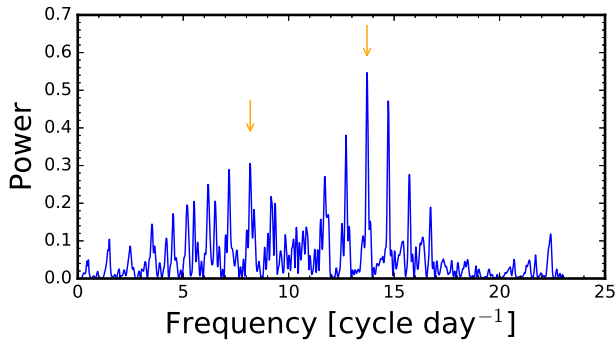}
\caption{Power spectrum for superhumps. For calculating the power spectrum using Lomb-Scargle Periodogram, $BVI$ data with 2 days before and 12 days after the superoutburst peak are combined. The local maximum frequencies of power are 7.18 and 13.72 cycle day$^{-1}$, which are shown as orange arrows, respectively. \label{fig09}}
\end{figure}


\subsection{\ylee{Periodic humps}} \label{ssec:hump}


In order to investigate periodic humps in the super/long outburst of {\target}, we calculate $\Delta$mag (observation $-$ best-polynomial fit) between 2 days before and 12 days after the superoutburst peak  and combine $BVI$-band data to increase the sample points. Then we conduct Lomb-Scargle analysis\footnote{astropy.stats.LombScargle is implemented \citep[\url{http://docs.astropy.org/en/stable/stats/lombscargle.html};][]{Ast13,Ast18} }. The resultant power spectrum is shown in Figure~\ref{fig09}.  There are possibly two local peaks. The local maximum powers are at frequencies of 7.18 and 13.72  ${\rm cycle}\,{\rm day}^{-1}$ (orange arrows). Uncertainties of these frequencies are roughly 0.01 ${\rm cycle}\,{\rm day}^{-1}$ \citep{Pat03} and the false alarm probabilities of the powers are 9.187$\times10^{-8}$ and 1.107$\times10^{-17}$.

\begin{figure}
\epsscale{1.25}
\plotone{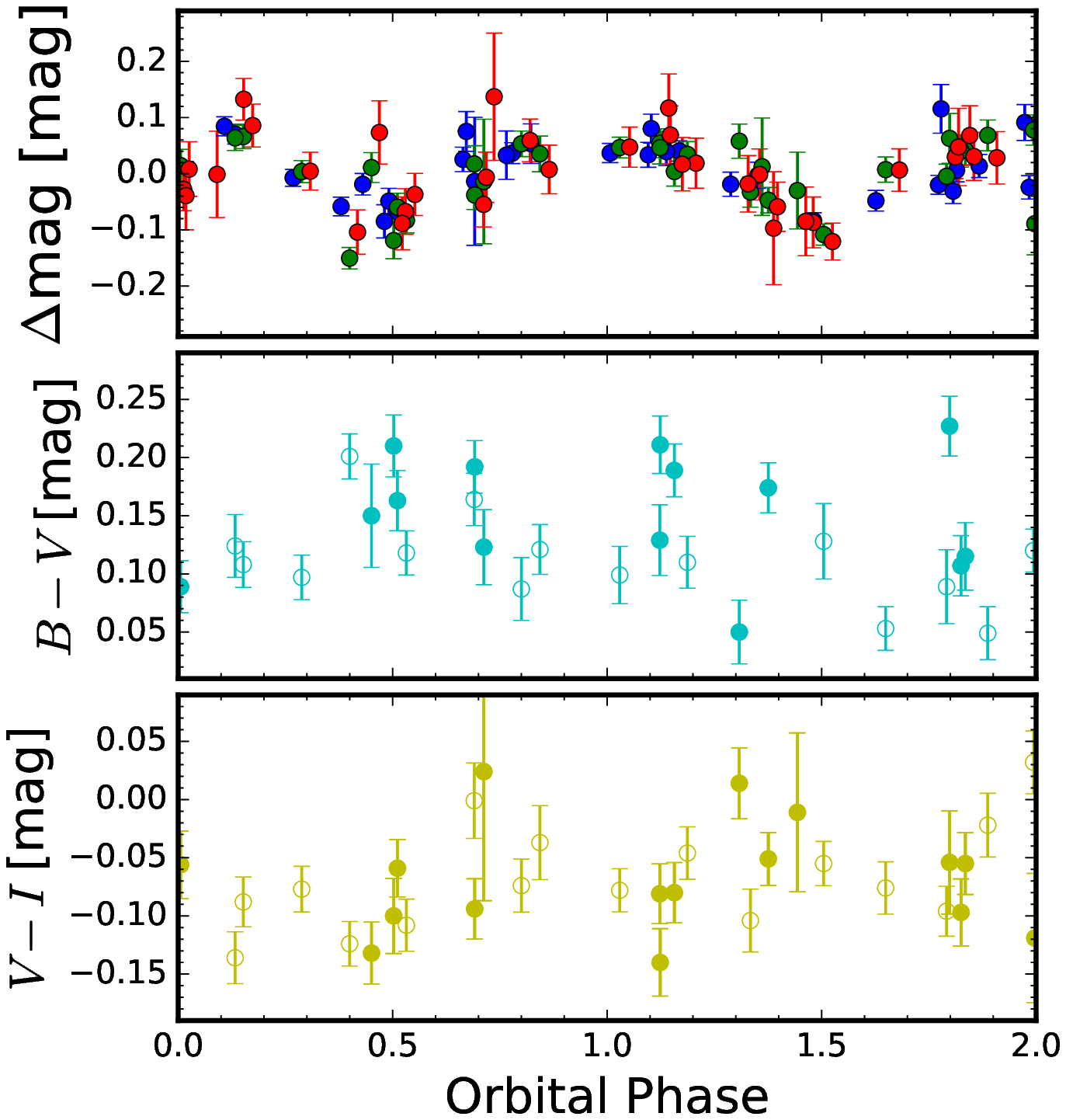}
\caption{(top) Phase diagram for the superhumps of {\target}. The phases are folded with the maximum frequency of 13.72 cycle day$^{-1}$. $BVI$ bands are shown as blue, green, and red circles, respectively. (middle) $B-V$ color. (bottom) $V-I$ color as a function of orbital phase. In the middle and bottom panels, open circles are data for between 2 days before and 3 days after the superoutburst peak. The filled circles are data for between 4 days and 12 days after the superoutburst peak. \label{fig10}}
\end{figure}

Based on the maximum frequency (13.72  ${\rm cycle}\,{\rm day}^{-1}$), we show a phase diagram in the top panel of Figure~\ref{fig10}. The periodic variations of $\Delta$mag are clearly shown. The existence of the periodicity is also shown in each band (blue circles for $B$, green circles for $V$, red circles for $I$), which strengthens that the periodicity is real and there is the existence of \ylee{periodic} humps. In the case of the second local maximum frequency, it shows a similar phase diagram with that of the maximum frequency. 


In the middle and bottom panels of Figure~\ref{fig10}, $B-V$ and $V-I$ colors of the \ylee{periodic humps} change as a function of orbital phase. Both colors have red peaks around 0.5 and 1.5 orbital phases showing the brightest peak of \ylee{periodic humps}. When \ylee{periodic humps} have the faintest peak (orbital phase $\sim$ 1.0), the colors are blue. \ylee{Two different mechanisms have been proposed to explain periodic color variations in dwarf nova humps. The first one is the color change caused by changing visibility of a hot spot on the accretion disk to explain the color changes observed from short orbital period ($<$2 hours) SU UMa types or long orbital period ($>$3 hours) U Gem types \citep{Pat81,Sma05}. This mechanism predicts bluer color at the maximum brightness of periodic humps than that at the minimum brightness, which appears to be inconsistent with our observations of {\target} (see Figure~\ref{fig10}). The other mechanism, on the other hand, is based on color change caused by the expansion of a low temperature regions, which predicts red colors at the maximum brightness of superhumps \citep{War95,Mat09,Neu17,Shu18}. The color variations observed from {\target} are consistent with prediction by this mechanism, which is suggestive that {\target} is an SU UMa type dwarf nova. We note, however, since there still exists a significant uncertainty in our understanding of the color variation in period humps, we can not rule out the first mechanism \citep{Mat09}.}

When we divide the data into two groups being around 0.5 days (between $-$2 days and 3 days) and 5.5 days (between 4 days and 12 days) after the \ylee{super/long} outburst peak in the middle panel of Figure~\ref{fig03}, 0.5 day region (open circles) has systematically blue color in $B-V$ than the 5.5 day region (filled circles) in the middle panel of Figure~\ref{fig10}. This implies that we observed the red colors of the \ylee{periodic} hump variations by chance in the 5.5 day region.

\begin{table*} 
\renewcommand{\thetable}{\arabic{table}}
\centering
\caption{Parameters of {\target}} \label{tab04}
\begin{tabular}{c c }
\tablewidth{0pt}
\hline
\hline
Parameter & Values (for max. freq. / local second max. freq.)  \\
\hline
\decimals
Expected type of dwarf nova    &  \ylee{SU UMa} / \ylee{U Gem}        \\
\ylee{Frequency of humps} ($f_s$)   &  13.72 $\pm$ 0.01 / 7.18 $\pm$ 0.01 cycle day$^{-1}$        \\
\ylee{Period of humps} ($P_s$)   &  1.750 $\pm$ 0.001 / 3.343 $\pm$ 0.004 hour        \\
Orbital period$^a$ ($P_o$) &  1.69 $\pm$ 0.01 / 3.343$^g$ $\pm$ 0.004 hour   \\
Mass of primary star$^{b}$ ($M_1$)   &  0.75 / 0.75 $M_{\odot}$     \\
Mass of secondary star$^c$ ($M_2$)  &  0.11 $\pm$ 0.01 / 0.21 $\pm$ 0.01 $M_{\odot}$      \\
$I$-band absolute magnitude of secondary ($M_{I2}$)  &  11.56 $\pm$ 0.01 / 9.33 $\pm$ 0.01 mag      \\
Effective temperature of secondary star$^d$ ($T_{eff2}$)     &  2976 $\pm$ 10 / 3319 $\pm$ 10 K  \\
Absolute magnitude of superoutburst peak$^e$ ($M_{p}$)  &  4.61 $\pm$ 0.41 / 4.29 $\pm$ 0.41 mag    \\
Distance \ylee{modulus}$^f$ ($m-M$)   &  14.33 $\pm$ 0.41 / 14.66 $\pm$ 0.41 mag      \\
Distance from the Sun ($d_S$)   & 7.3$^{+1.5}_{-1.3}$ / 8.5$^{+1.8}_{-1.5}$ kpc    \\
Distance from the Galactic center ($d_{GC}$)   & 13.8$^{+1.3}_{-1.1}$ / 14.9$^{+1.6}_{-1.3}$ kpc   \\
Scale height from the Galactic plane (Z)       & $-$1.7$^{+0.3}_{-0.4}$ / $-$2.0$^{+0.3}_{-0.4}$ kpc    \\

\hline

\multicolumn{2}{l}{$^a$ Equation (7) of \citet{Men99}} \\
\multicolumn{2}{l}{$^b$ Table 8 of \citet{Kni11} \ylee{only provides $M_1 = 0.75$\, $M_{\odot}$.} } \\
\multicolumn{2}{l}{$^{c,d}$ Table 6 of \citet{Kni11}; Uncertainties are propagated from that of $P_o$} \\
\multicolumn{2}{l}{$^e$ Equation (4) of \citet{Pat11}} \\
\multicolumn{2}{l}{$^f$ for \ylee{super/long outburst}} \\
\multicolumn{2}{l}{$^g$ \ylee{In the case of U Gem types, $P_o$ is the same with period of humps.}} \\
\multicolumn{2}{l}{$^*$ \ylee{It is noted that the study of \citet{Kni11} was stemed from dwarf novae with solar metallicities.} } \\
\end{tabular}
\end{table*}

\subsection{Distance and parameters of {\target}} \label{ssec:pars}

In SU UMa-type dwarf novae, the superhump period ({\Ps} $ = 1/f_s$) tightly correlates with an orbital period ({\Po}), where $f_s$ is a superhump frequency \citep{Men99}. \ylee{In U Gem-type dwarf novae, the hump period is considered as an orbital period \citep{Rut92,Web99,Bap00}}. Orbital periods of the maximum and the second local maximum power are 0.0705 \ylee{days (1.69 hours) calculated by the equation (7) of \citet{Men99} and 0.1393 days (3.34 hours), respectively.} Parameters of dwarf novae including the masses of primary and secondary, the effective temperature and absolute magnitude of the secondary, and the absolute magnitude of the \ylee{super/long} outburst peak can be derived from the orbital period \citep{Beu98,Men99,Kni11,Har16,Otu16}. To derive the absolute magnitude of the \ylee{super/long} outburst and the other parameters, we use the equation (4) of \citet{Pat11} and Tables of \citet{Kni11}. The derived parameters of {\target} shown in Table~\ref{tab04}.

\ylee{We estimate the distance to {\target} by comparing the observed $I$-band magnitude in quiescence with he known absolute magnitudes of the secondary in dwarf novae \citep{Kni11}. Considering that the observed $I$-band magnitude also has contributions from the accretion disk and hot spot \citep{Szk76,Mat09}, this estimation gives only a lower limit of the distance. For the two cases of the orbital periods of 1.69 and 3.34 hours, we obtain the distance of 1.7 and 4.7 kpc, respectively. Since the absolute magnitudes of the secondary in \citet{Kni11} are for those with solar metallicities, the distance of {\target} can be increased to 2.7 and 7.4 kpc if it is with lower metallicity. For low metallicities, we consider that the K- or M-type secondaries with low metallicities ($Z\lesssim0.001$) may have about 1 mag brighter than those with solar metallicities \citep{Yi01,Her15}.}

In addition to quiescent magnitudes, in order to derive a distance to {\target}, we use the \ylee{super/long} outburst peak magnitude. The derived absolute magnitude combined with the observed apparent magnitude of the \ylee{super/long} outburst peak provides the distance ($d_{S}$) of $7.3_{-1.3}^{+1.5}$/$8.5_{-1.5}^{+1.8}$ kpc from the Sun.

Using the Galactic coordinates ($l$,$b$) of {\target} and assuming the distance of the Sun from the galactic center to be $R_0 = 8.3$ kpc \citep{Rei14}, with the orbital period of 1.69 hours, we derived a Galactocentric distance ($\sqrt{d_S^2 + R_0^2 - 2R_0d_Scos(l)cos(b)} = 13.8_{-1.1}^{+1.3}$ kpc) and a height ($d_Ssin(b) = -1.7_{-0.4}^{+0.3}$ kpc) from the Galactic plane. In the case of the orbital period of 3.34 hours, the Galactocentric distance is $14.9_{-1.3}^{+1.6}$ kpc and the height from the Galactic plane is $-2.0_{-0.4}^{+0.3}$ kpc.

\section{Discussion} \label{sec:discuss}

\subsection{Outbursts} \label{disc:outburst}

\ylee{For the outbursts of dwarf novae, two types of outburst mechanisms have been suggested \citep{Sma84b,Can86,Osa96,Fra02,Sch07}. The first one is an inside-out outburst where matter is accreted from the secondary and thermal instability is started at the inner disk. Then, the thermal instability propagates to the outer disk. In contrast to the inside-out outburst, in the outside-in outburst,  thermal instability is triggered at outer disk first and the instability propagates inward. In {\target}, the asymmetric shape of the super/long outburst is suggestive of outside-in accretion disk instability \citep{Sma84b,Can86,Can87}, although we cannot completely rule out the possibility of inside-out origin if the accretion disk is affected by stream impact from the secondary and tidal torque dissipation \citep{Bua01}. Although the light curves of the normal/short outburst (Figure~\ref{fig04}) also  indicate that they are asymmetric, the incomplete sampling makes it difficult to reach a firm conclusion whether or not it is indeed asymmetric.}

\ylee{The two outbursts show distinctively different color evolution: while the super/long outburst shows no color variation in $B-V$ and $V-I$ during the outburst, the normal/short outburst shows blue and red evolution in $B-V$ and $V-I$ color, respectively, during the outburst. The invariant colors are consistent with expectation of constant viscosity model in the accretion disk during the outburst, suggesting constant mass flow rate throughout the accretion disk \citep{Can87}. It is difficult, on the other hand, to explain the observed color evolution during the normal outburst where the $B-V$ and $V-I$ colors evolve in opposite directions: the former becomes bluer, while the latter become redder (Figure~\ref{fig03} and \ref{fig04}). Similar patterns have been observed in $B-V$ and $V-R$ of RZ LMi \citep{Shu18}. However, the origin of the color evolution is unknown. We also note that such color evolution is observed in the classical nova, where they suggest that the formation of a dust shell may have an additional thermal emission component, although how a dust shell can be formed is not explained \citep{Ant17}. Here, for {\target}, the color evolution during the normal outburst may be explained if there is new cold material accreted from the secondary at the outer part of the accretion disk, increasing the flux in the $I$-band while overall temperature of the accretion disk increases during the outburst \citep{Osa89}. More detailed analysis is required to investigate this possibility more thoroughly.}

\subsection{\ylee{Halo or thick disk} dwarf novae}

In the Milky Way, PopI showing young age, high metallicity, and cold kinematics is located in the thin disk, whereas PopII characterized by old age, low metallicity, and hot kinematics is classified as a \ylee{halo or thick disk} components \citep{Lee11,Li18}. PopI is generally concentrated in the Galactic disk (height of $\lesssim$0.6 kpc) and PopII is populated on further distance (height of $\gtrsim$1 kpc) \citep{Lee11,Mat18}. In the same situation, dwarf novae follow the same trend. That is, PopI dwarf novae are more concentrated in the Galactic disk than PopII dwarf novae \citep{How90,Ak13,Uth11,Ozd15}. Since PopI dwarf novae are located nearby the Sun, they have been found and well studied \citep{Cop16,Otu16}. In the case of PopII, although there have been attempts to find distant dwarf novae (possible PopII) in \ylee{halo or thick disk} \citep{Haw87,How90,Sha03}, few candidates have been found because they reside at large distance from the Sun. \ylee{Notably, no distance ($>$5 kpc from the Sun) PopII dwarf nova has been observed with light curves other than the nearby ($250$ pc) source SDSSJ1507+52 \citep{Uth11}}.

{\target} is a possible candidate of PopII dwarf nova with $BVI$ light curves and the faintest quiescent magnitudes that have ever observed \citep[cf.][]{Cop16}. {\target} has a Galactocentric distance of 13.8 kpc and a height of $-$1.7 kpc. The large height of {\target} safely rejects the possibility that {\target} is a thin disk component. Considering the Galactic position of {\target}, it is located near the Perseus arm which is warped toward the direction of {\target} \citep{Koo17}. However, the warped disk reaches a height of about $-$1 kpc \citep{Koo17,Sko18}. This fact implies that {\target} is well separated from the warped disk and may be \ylee{halo or thick disk} populations \citep{Ozd15,Mat18}.

Theoretically, dwarf novae with the low-metallicity secondary expect to have a brighter accretion disk, owing to larger accretion rate, than dwarf novae with the solar-metallicity secondary \citep{Ste97}. In observation, 8 dwarf novae in a globular cluster, 47 Tucanae, have fainter accretion disk luminosities than those of field dwarf novae \ylee{with solar metallicities} \citep{Edm03}. This inconsistency between theory and observation may stem from a poor understanding of the emission mechanism of the accretion disk related to metallicity. It is worth comparing dwarf novae with low- and high-metallicity secondaries to tune the emission mechanisms.

\begin{figure}
\epsscale{1.25}
\plotone{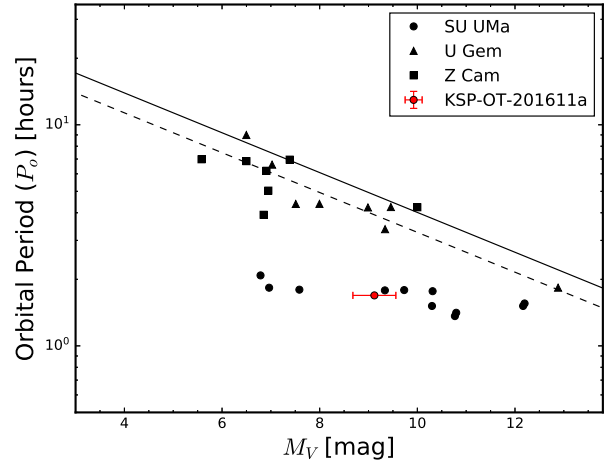}
\caption{Orbital periods ($P_o$) as a function of $V$-band absolute magnitude in quiescence ($M_V$). Black symbols are various types of dwarf novae. The solid line is for the secondaries with solar metallicity \citep{War95}. The dashed line is for the secondaries with low-metallicity ($Z \lesssim 0.001$) which is 1 mag brighter than the solid line. {\target} is shown as the red circle. \label{fig11}}
\end{figure}

\ylee{In Figure~\ref{fig11}, we compare the $V$-band absolute magnitudes in quiescence ($M_V$; red circle) of {\target} with those of solar-metallicity secondaries \citep[solid line;][]{War95} and those of low-metallicity secondaries (dashed line; $Z \lesssim 0.001$) being 1 magnitude brighter than the solid line. At a given orbital period, $M_V$ of {\target} is $\sim$4 mag and $\sim$5 mag brighter than the brightness of the secondaries with solar and low metallicities, respectively. This means that {\target} has brighter accretion disk than the secondaries with solar and low metallicities. At a given orbital period, SU UMa types of field dwarf novae \citep[PopI; black circles][]{Ver97} also have brighter accretion disks and have similar $M_V$ of {\target}. Although solar-metallicity dwarf novae have fainter accretion disks than low-metallicity dwarf novae in the theoretical model, our results show that solar-metallicity dwarf novae and {\target} have similar accretion disks. This may be interpreted as that metallicity effects on accretion disk luminosities are not strong \citep{Bel18}.}

Discovering more PopII dwarf novae in a large distance (Galactic height of $\gtrsim$1 kpc) gives a chance to modify the emission mechanisms in the accretion disk. Although this rare distant dwarf nova is difficult to be detected, KSP is an optimal survey to find distant PopII dwarf novae. For example, KSP-OT-201503a, located at $(l,b)_{J2000}=(252.04513^{\circ}, 19.77151^{\circ})$ and having a distance of $\sim$$4.0$ kpc from the Sun and a height of 1.4 kpc from the Galactic plane \citep{Bro18}, is also a possible PopII dwarf nova observed in KSP.

\section{Summary and conclusions} \label{sec:conclusion}

In this paper, we present our discovery and multi-color photometric monitoring of the new dwarf nova {\target} observed in October 2016--May 2017. Below we provide a summary of our study and conclusions.

\begin{itemize}
    \item We detect two outbursts of \ylee{possible} SU UMa-type dwarf nova separated by $\sim$91 days. The first outburst (superoutburst) lasts about 18 days with the peak outburst amplitude of $\sim$4.51 mag in the $V$ band, while the second one (normal outburst) remains only 6 days with a slightly smaller outburst amplitude of $\sim$4.05 mag in the $V$ band. Both outbursts show an asymmetric shape in light curve evolution with a rapid pre-peak ascent followed by a slow post-peak decline. 

    \item The two outbursts show a clear difference in their colors: while the superoutburst shows no color evolution during the outburst in $B-V$ and $V-I$, the normal outburst shows that $B-V$ moves in the blue direction and $V-I$ is reddened during the outburst. \ylee{This may result from the difference of time when the cooling wave is formed in the accretion disk.}

    \item The observed light curves of the superoutburst are consistent with the presence of superhumps of 1.69-hour period superimposed on the outburst. The superhumps become redder as they become brighter, showing that dwarf novae reach the superhump maximum by the expansion of a low-temperature region in the accretion disk. 
    
    \item The observed magnitude of the source during its quiescent phase is $V \simeq 23.45$ mag with $B-V$ and $V-I$ color of 0.07 and 0.76 mag which cannot be explained by a typical binary system for a dwarf nova composed of a white dwarf and a main-sequence star. This strongly indicates the presence of substantial emission from the accretion disk during the quiescent phase.

    \item The estimated Galactocentric distance of {\target} is 13.8 kpc, making it one of the most distant dwarf novae ever observed. This also gives 1.7 kpc as a height of the source from the Galatic plane. It is, therefore, highly likely that the source is a rare PopII dwarf nova in the \ylee{halo or thick disk} of the Galaxy. The more detailed observations and studies of the source may provide a new clue to understanding how the metallicity is related to the accretion disk process. 

\end{itemize}

\acknowledgments

This research has made use of the KMTNet system operated by the Korea Astronomy and Space Science Institute (KASI) and the data were obtained at three host sites of CTIO in Chile, SAAO in South Africa, and SSO in Australia.

\ylee{\software{DAOPHOT \citep{Ste87}, astropy \citep{Ast13,Ast18}}}
\ylee{\facility{KMTNet}}

\bibliographystyle{apj}
\bibliography{n2292tran}

\end{document}